\documentstyle[epsfig]{IOPConf}
\begin{document}
\frenchspacing
\newcommand{\bm}{\bibitem}
\newcommand{\ud}{\bf}
\renewcommand{\thefootnote}{\fnsymbol{footnote}}
\title{Breakup of $^8$B and the 
$^7$Be($p,\gamma)^8$B reaction\footnote{Work supported by EPSRC,
UK, grant nos. J/95867 and L/94574}}
\longauthor{ R. Shyam\dag ~ and I.J. Thompson\ddag}
	{R. Shyam and I.J. Thompson}
\address{\dag Saha Institute of Nuclear Physics, Calcutta, India.\\
\ddag  Department of Physics, University of Surrey, Guildford,
Surrey GU2 5XH, U.K.}

\beginabstract
The calculated rate of events in some of the existing solar neutrino
detectors is directly proportional to the rate of the 
$^7$Be($p,\gamma)^8$B
reaction measured in the laboratory at
low energies. However, the low-energy cross sections of this 
reaction are quite uncertain as various measurements differ from each 
other by 30-40 \%. The Coulomb dissociation process which reverses 
the radiative capture by the dissociation of $^8$B in the 
Coulomb field of a target, provides an alternate way of accessing 
this reaction. While this method has several advantages
(like large breakup cross sections and flexibility in the kinematics),
the difficulties arise from the possible interference by the 
nuclear interactions, uncertainties in the contributions of the
various multipoles and the higher order effects, which should be  
considered carefully. We review the progress made so far in 
the experimental measurements and theoretical analysis of the
breakup of $^8$B and discuss the current status of the low-energy
cross sections (or the astrophysical $S$-factor) of the   
$^7$Be($p,\gamma)^8$B reaction extracted therefrom.  
The future directions of the experimental and theoretical
investigations are also suggested.
\endabstract

 

\section{Introduction}
The $^8$B isotope produced in the Sun via the radiative capture reaction
$^7$Be(p,$\gamma$)$^8$B
is the principal source of the 
high energy neutrinos detected in the Super-Kamiokande (SK) and $^{37}$Cl
detectors~\cite{bahc89}. In fact the calculated rate of events in 
SK as well as SNO detectors~\cite{bahc98}  
is directly proportional to the rate of this reaction measured 
in the laboratory at low energies ($\sim$ 20 keV)
\cite{bahc98}. Unfortunately, the measured
cross sections (at relative energies ($E_{CM}$) of 
[$\mbox{p} - {^7}\mbox{Be}$] $>$ 200 keV)
disagree in absolute magnitude and the value extracted by extrapolating
the data in the region of 20 keV differ from each other by 30-40 $\%$.
This makes the rate of the reaction $^7$Be($p,\gamma)^8$B the most poorly
known quantity in the entire nucleosynthesis chain leading to
the formation of $^8$B~\cite{adel98}. It may be noted that 
the rate of the
$^7$Be(p,$\gamma$)$^8$B reaction is usually given in terms of the 
zero-energy astrophysical $S$-factor, $S_{17}(0)$. 

The Coulomb dissociation (CD) method provides an alternative indirect
way to determine the cross sections for the radiative capture reactions
at low energies~\cite{baur94,baur97,shya96,shya97,shya99}.
In this procedure it
is assumed that the break-up reaction
$a+ Z \rightarrow (b + x) + Z$ proceeds entirely via
the electromagnetic interaction; the two
nuclei $a$ and $Z$ do not interact strongly.
By further assuming that the electromagnetic excitation
process is of first order, one can relate directly
(see {\it e.g.} Refs.~\cite{baur94,baur97}) the
measured cross-sections of this reaction to those of the
radiative capture reaction $b + x \rightarrow a + \gamma $.
Thus, the astrophysical S-factors of the radiative capture processes
can be determined from the study of break-up reactions under
these conditions.

However, in the CD of $^8$B, the contributions of
$E2$ and $M1$ multipolarities  can be
disproportionately enhanced in certain kinematical
regimes~\cite{lang94,gai95}. Furthermore, interference from the
nuclear breakup processes may also be considerable in some
regions. Therefore, a careful
investigation~\cite{shya99,gai98} is necessary to isolate the
conditions in which these terms have negligible effect on the
calculated breakup cross sections.

Motobayshi et al.~\cite{moto94} have performed the first 
measurements (to be referred as RIKEN-I) of the 
dissociation of $^8$B into the $^7$Be$-p$ low
energy continuum in the field of $^{208}$Pb with a radioactive
$^8$B beam of 46.5 MeV/nucleon energy. Assuming a pure E1
excitation, the Monte Carlo simulation of their data predicts a
$S_{17}(0)=16.7 \pm 3.2$ eV barn, which is considerably
lower than the value of 22.4 $\pm$2.0 eV barn used by
Bahcall and Pinsonneault~\cite{bahc95} in
their standard solar model (SSM) calculations. This generated 
a lot of interest in the studies of the breakup reactions of
$^8$B. Since, under the kinematical conditions of the RIKEN-I 
experiment the $E2$ component of breakup may be disproportionately
enhanced, attempts were made to determine this component by
extending the angular range of the measurements in the RIKEN-I
data in a repeat experiment~\cite{kiku97}
(to be referred as RIKEN-II) to larger 
angles which are expected to be more sensitive to this multipolarity.
On the other hand, measurements of the breakup of $^8$B were also
carried out at subCoulomb beam energies~\cite{joha96}
where $E2$ multipolarity is expected to dominate according to
the semi-classical theory of the Coulomb excitation~\cite{ald75}.
Measurements of the breakup of $^8$B have also been performed at
the relativistic energies of 250 MeV/nucleon at GSI Darmstadt.

In this review, we present the latest status of the analysis of the 
available experimental data on the breakup of $^8$B and of the
extracted $S_{17}$ value therefrom. Results obtained from
the both semiclassical and full quantum mechanical calculations
are discussed in the next two sections. Conclusions and the outlook
in presented in section 4.

\section{Semiclassical calculations}
\subsection{RIKEN data, $E_{beam} \sim$ 50 MeV/nucleon}

An analysis of the RIKEN-I data was presented in~\cite{shya96},  
where the breakup cross sections of $^8$B corresponding to 
$E1$, $E2$ and $M1$ multipolarities were calculated within a semiclassical
theory of Coulomb excitation, which included simultaneously 
the effects of Coulomb recoil and relativistic retardation. This was 
achieved by solving the general classical problem of the motion of two  
relativistic charged particles~\cite{alex89}.   
The role of the nuclear excitations was also investigated by 
performing full quantum mechanical calculation of the
Coulomb, nuclear as well as of their interference terms, using  
a collective model prescription for the nuclear
form factor. It was found that 
nuclear effects modify the pure Coulomb amplitudes
very marginally in the entire kinematical regime of the RIKEN-I data. 

The double differential cross-section for the Coulomb excitation
of a projectile from its ground state to the continuum, with a
definite multipolarity of order $\pi\lambda$ is given
by~\cite{baur94,baur97,shya96}   
\begin{eqnarray} 
\frac{d^2 \sigma}{d \Omega dE_{\gamma}} & = & \sum_{\pi\lambda}
                                     \frac{1}{E_\gamma}
                       \frac{dn_{\pi \lambda}}{d \Omega} \sigma_{\gamma}^{\pi
                              \lambda}(E_\gamma),
\end{eqnarray}
where
$\sigma_{\gamma}^{\pi \lambda}(E_{\gamma})$ is the cross-section for
the photodisintegration process $ \gamma + a \rightarrow b + x$,
with photon energy $E_\gamma$, and multipolarity
$\pi\,=\,$ E (electric) or M (magnetic), and 
$\lambda\,=\, 1,2...$ (order), which is related to that of
the radiative capture process $\sigma (b + x
\rightarrow a + \gamma)$ through the theorem of detailed
balance. In terms of the astrophysical S-factor, $S(E_{cm})$, 
we can write 
\begin{eqnarray}
\sigma (b + x \rightarrow a + \gamma ) & = & \frac{S(E_{cm})}{E_{cm}}
                                      \exp(-2 \pi \eta(E_{cm})),
\end{eqnarray}
where $\eta\, =\,\frac{Z_b Z_x e^2}{\hbar v}$,
with $v$, $Z_b$ and $Z_x$ being the relative center of mass
velocity, and charges of the fragments $b$ and $x$ respectively.

In most cases, only one or two multipolarities dominate the 
radiative capture as well as the Coulomb dissociation cross sections.
In Eq. (1)
$n_{\pi\lambda}(E_\gamma)$ represents the number of equivalent
(virtual) photons provided by the
Coulomb field of the target to the projectile, which is  
calculated by the method discussed in Ref.~\cite{shya96,alex89} .   
$S(E_{cm})$, can be 
directly determined from the measured Coulomb dissociation
cross-sections using Eqs. 1 and 2.

In Fig. 1, we show the comparison of the calculated~\cite{shya96}
Coulomb dissociation 
double differential cross sections with the corresponding data of
Ref.~\cite{moto94}   
as a function of the scattering angle $\theta_{cm}$ of the
excited $^8$B (center of mass of the $^7$Be$ + p $ system) for
three values of the $E_{cm}$.
The calculated $E1$, $E2$ and $M1$ cross sections 
are folded with an efficiency matrix provided to us by the RIKEN group.
The solid lines in Fig. 1 show the calculated $E1$ cross sections  
obtained with S-factors $(S_{17})$ that provide
best fit to the data (determined by $\chi ^2 $ minimization procedure).
These are $(17.58 \pm 2.26)$ eV barn, $(14.07 \pm 2.67)$
eV barn  and $(15.59 \pm 3.49)$ eV barn at $E_{cm}$= 0.6 MeV,
0.8 MeV and 1.0 MeV respectively. 
By using a direct extrapolation procedure, the best fit
``$E1$ only'' $S_{17}$ factors,  
give a $S_{17}(0) \,= \,(15.5 \pm 2.80)$ eV barn.
 
The contributions of the $E2$ and $M1$ excitations
are calculated by using the radiative capture cross sections 
$\sigma(p+^7$Be$\rightarrow ^8$B$ +\gamma)$, given by the models of
\begin{figure}[htb]
\begin{center}
\epsfig{file=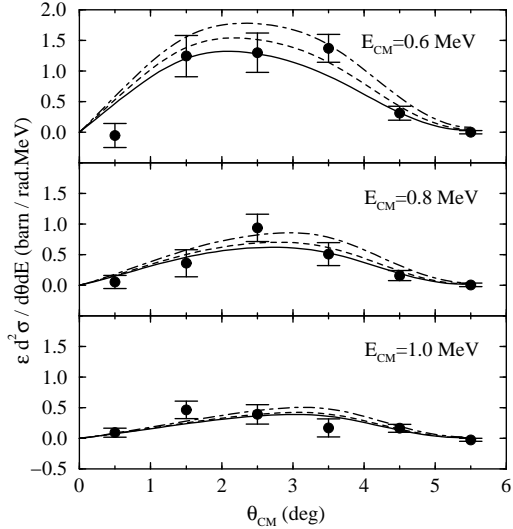,height=7.0cm }
\end{center}
\caption[C1]{{\captionfont
Comparison of experimental and theoretical 
Coulomb  dissociation yields (cross section $\times$ detector
efficiency) as a function of $\theta_{cm}$ for the $E_{cm}$ 
values of 0.6 MeV, 0.8 MeV and 1.0 MeV.
Solid lines show the calculated pure $E1$ Coulomb dissociation
cross sections obtained with best fit 
values of S factors as discussed in the text. The dashed and dashed dotted
curves represent the sum of $E1$, $E2$ and $M1$ contributions with 
latter two components calculated with capture cross sections 
given in the models of TB~\protect\cite{type94} 
and KPK~\protect\cite{kimp87} respectively.
The experimental data is taken from Ref.~\protect\cite{moto94}.}}
\label{fig:figa}
\end{figure}
\noindent
{}Typel and Baur (TB)~\cite{type94} and Kim, Park and Kim (KPK)~\cite{kimp87}.
We have used as input the corresponding S factors averaged 
over energy bins of experimental uncertainty in the
relative energy of the fragments. In Fig. 1, the dashed (dashed
dotted) line shows the $E1$ (with best fit
$S_{17}$) + $E2$ + $M1$ cross sections, with $E2$ and $M1$ components 
calculated  with TB (KPK) capture cross sections. It is clear that
the magnitude of the $E2$ contributions to the RIKEN-I data 
depend significantly on the nuclear structure model used to
calculate the corresponding capture cross sections, and it  
is difficult to draw any definite conclusion about the extent of
its role in the RIKEN-I data from this analysis. The $M1$ component
contributes insignificantly and unlike the $E2$ component
it not as model dependent. 

Since at larger scattering angles, the angular distributions of the
Coulomb breakup of $^8$B are expected to be more sensitive to the
$E2$ component, the RIKEN group has repeated their
experiment~\cite{kiku97} 
where the angular range of the data was extended up to 9$^\circ$.
In Fig. 2, we show a comparison of the calculated Coulomb dissociation
cross sections for the double differential cross sections 
with the RIKEN-II data. In these calculations the capture cross 
sections have been taken from Esbensen and Bertsch~\cite{esbe96},
which predicts a $S_{17}(0)$ = 18.5 ev barn. Since we have not
used any arbitrary normalization constant in the theoretical calculations
shown in this figure,  RIKEN-II data seems to be  
consistent with a slightly larger value of $S_{17}(0)$ as compared to 
RIKEN-I. We also note that while the $E2$ component contributes significantly
to the total cross sections at all the angles in the energy bin 2000-2250
keV, it is dominant beyond 6$^\circ$ in the lower energy range of
$E_{CM}$. However, at
larger angles the nuclear breakup effects are also expected to be more
important. Therefore, it would be necessary to include these effects 
before drawing any conclusion about the role of $E2$ multipolarity
in this data. 

In Ref.~\cite{kiku97}, an analysis of the data was performed
within the distorted wave Born-approximation 
including the nuclear effects, where it was  
concluded that the $E2$ component and the nuclear
breakup effects are considerably smaller. However, they  
use a collective model prescription to calculate the inelastic
nuclear form factor (see eg.~\cite{shya96}).  Due
to a long tail in the $^8$B g.s wave function this procedure is
unlikely to be accurate. Furthermore, Coulomb breakup is calculated
by a point-like projectile approximation (PLPA) in these studies (and also
in the semiclassical calculations presented above), and its
range of validity is yet to be determined for this projectile.

It is therefore, necessary to perform a full quantal mechanical 
analysis of the RIKEN-II data in order to check the validity of 
various assumptions of the Coulomb dissociation method. This will
be presented in section \ref{quansect}.

\begin{figure}[htb]
\begin{center}
\mbox{\epsfig{file=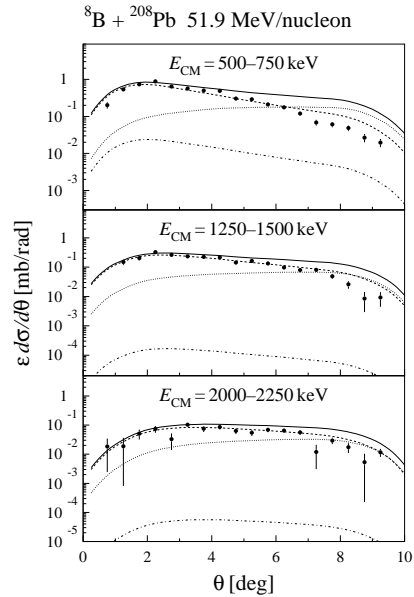,height=8.0cm }}
\end{center}
\caption[C2]{{\captionfont $E1$ (dashed line), $E2$ (dotted line), and $M1$
(dashed-dotted line) components of the Coulomb dissociation 
cross section $\epsilon d\sigma/d\theta$ as a function of 
the scattering angle in the dissociation of $^8$B on $^{208}$Pb
target at the beam energy of 51.9 MeV/nucleon. The solid line
shows their sum. Results for relative energy bins of (a) 500-750 keV,
(b) 1250-1500 keV and (c) 2000-2250 keV are shown. $\epsilon$ is 
the detector efficiency. 
The experimental data and $\epsilon$ are taken from
Ref.~\protect\cite{kiku97}.}}
\label{fig:figb}
\end{figure}

\subsection{Notre Dame data, $E_{beam}$ = 25.8 MeV}
 
The Notre Dame group has measured the breakup of $^8$B on the
$^{58}$Ni target at the beam energy of 25.8 MeV, well below the
Coulomb barrier, where the $E2$ component is expected to dominate
the CD process~\cite{joha96}. However, the reliable
extraction of the $E2$ component from this data, where only the
the integrated cross section of the $^7$Be fragment is measured,
is still doubtful. The analysis of the data reported in Ref.~\cite{joha96}
used the Alder-Winter's semiclassical theory of Coulomb excitation,
where the final state is treated as a two-body
system, thus assuming that the measured angles of $^7$Be were
equal to those of the $^7$Be-$p$ center of mass. The inadequacy
of this assumption has been demonstrated in~\cite{shya97}.
Furthermore, the total breakup cross section reported in this
experiment could not be reproduced  within the Alder-Winther
theory even if a wide variety of structure models of $^8$B were
used~\cite{nune98a}. Therefore, the uncertainty about the magnitude
of the $E2$ cross section calculated with various structure models
of $^8$B is not eliminated by the Notre Dame
measurements~\cite{joha96}.

Furthermore, the 
importance of the nuclear breakup effects in the kinematical
regime of the Notre Dame experiment has  been
emphasized in Ref.~\cite{nune98b}. Therefore, there is a need
to reanalyze this data using a quantum mechanical theory  
where the nuclear excitations and the three-body kinematics
are taken into account. 

\section{Full Quantum Mechanical calculations} 
\label{quansect}

A one-step prior-form DWBA analysis of the
$^8$B breakup data 
has been reported in~\cite{shya99} at both low and high energies
in order to check
the validity of various assumptions of the Coulomb dissociation
method. The breakup process is described as a single proton excitation
of the projectile from its ground state to a range of states in the
continuum, which is discretized by the method of continuum bins. 
Excitations to states corresponding to the relative energy
(of the $p-^7Be$ system) up to 3.0 MeV and relative partial
waves up to 3 have been taken into account.
The point-like projectile approximation as well
as collective model prescription for the nuclear form factor have 
been avoided,
by determining the nuclear and Coulomb parts by a single-folding 
method where the relevant fragment-target interactions are folded
by the projectile wave functions in the ground and continuum states.

\subsection{$^8$B Breakup at $\sim$ 50 MeV, RIKEN data}

In Fig. 3a, $E1$ and $E2$ components of the angular distributions
for the $^8$B + $^{208}$Pb $\rightarrow$ $^8$B$^*$ + $^{208}$Pb
reaction measured by the Kikuchi et al.~\cite{kiku97} at the beam
energy of 415 MeV are shown, for the pure Coulomb excitation case.
The dashed, dotted and solid lines represent $E1$, $E2$ and $E1+E2$
cross sections respectively which are obtained by the single-folding
procedure. Also shown in this figure are the corresponding results 
obtained by PLPA (curves with solid circles). We note that PLPA 
becomes inaccurate beyond 4$^\circ$ in this case.
Moreover, the $E2$ component of the pure Coulomb excitation becomes
increasingly important also after this angle.

In Fig. 3b, the cross sections obtained by summing
coherently the Coulomb and nuclear amplitudes (to be referred as {\it total}
in the following) are shown.  
The dashed and dotted lines show the dipole and quadrupole cross
sections respectively, while the solid line represents their sum. 
It can be noted that nuclear effects modify the pure Coulomb
$E1$ cross sections substantially after $\sim$ 4$^\circ$, and  
the $E2$ cross sections in the entire angular range.
However, since the $E2$ components are quite small at angles
$\leq$ 4$^\circ$, the difference between pure Coulomb and
{\it total} dipole + quadrupole cross sections is appreciable
only after this angle.

\begin{figure}[tb]
\begin{center}
\mbox{\epsfig{file=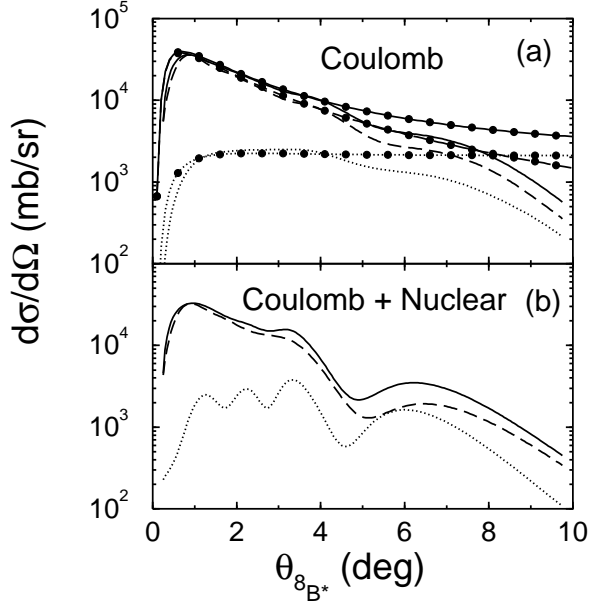,height=8.0cm}}
\end{center}
\vskip .1in
\caption{{\captionfont
Angular distribution for $^8$B+$^{208}$Pb $\rightarrow$ $^8$B$^*$($^7$Be+p)+
$^{208}$Pb reaction at the beam energy of 415 MeV.
(a) Results for pure Coulomb excitation, the dashed and
dotted curves represent the $E1$ and $E2$ cross sections while their sum is
depicted by the solid line. Also shown here are the results obtained with a
point-like projectile approximation (Alder-Winther theory), where
dashed and dotted lines with solid circles show the corresponding
$E1$ and $E2$
cross sections while the solid line with solid circles represents their
sum. (b) Coherent sum of Coulomb and Nuclear excitation calculations;
the dashed and dotted lines show the dipole and quadrupole components while
the solid line is their sum. }}
\label{fig:figc}
\end{figure}

\begin{figure}[tb]
\begin{center}
\mbox{\epsfig{file=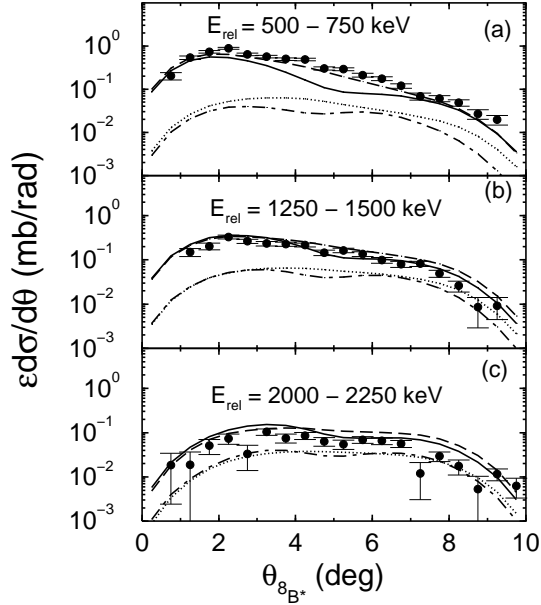,height=8.0cm}}
\end{center}
\caption{{\captionfont
Comparison of experimental and theoretical cross section
$\epsilon d\sigma/d\theta$ as a function of
the scattering angle $\theta_{8_{B^*}}$ 
for $^8$B+$^{208}$Pb $\rightarrow$ $^8$B$^*$($^7$Be+p)+
$^{208}$Pb reaction at the beam energy of 415 MeV.
Results for three relative energy bins of (a) 500-750 keV,
(b) 1250-1500 keV, (c) 2000-2250 keV are shown. $\epsilon$ is 
the detector efficiency. Solid lines show the calculated total Coulomb plus
nuclear dissociation cross sections while the dashed lines represents the
corresponding pure Coulomb dissociation result. Pure quadrupole  Coulomb and
Coulomb+nuclear cross sections are shown by dotted and dashed-dotted lines.
The experimental data and the detector efficiencies are taken
from~\protect\cite{kiku97}.} }
\label{fig:figd}
\end{figure}

Therefore, at RIKEN energies, the PLPA breaks down beyond 4$^\circ$,
where the Coulomb-nuclear interference effects as well as the quadrupole 
component of breakup is substantial. Hence, the Coulomb dissociation
method as used in e.g.  Ref.~\cite{shya96} to extract 
{}$S_{17}(0)$ from the measurements of the angular distributions in
the breakup of $^8$B on heavy target at RIKEN energies
($\sim$ 50 MeV/nucleon), is reliable only when data is taken at
angles below 4$^\circ$. 

In Figs. 4a, 4b and 4c the comparison of calculations~\cite{shya99}
for $\epsilon \cdot d\sigma/d\theta$ with the experimental data of
Kikuchi et al.~\cite{kiku97} is shown as a function of the scattering angle
$\theta_{8_{B^*}}$ of the excited $^8$B (center of mass of the $^7$Be+$p$
system) for three relative energy bins. The efficiency
($\epsilon$) matrix as well as angular and energy averaging were
the same as those   
discussed in Ref.~\cite{kiku97}.
The dashed and dotted lines are the pure Coulomb
$E1$+$E2$ and $E2$ cross sections respectively while the solid
and dashed lines are
the corresponding {\it total} cross sections. We note that the 
calculations are in fair agreement with the experimental data. 
No arbitrary normalization constant has been used in 
the results reported in this figure.  

The quadrupole component of breakup is significant at almost all the  
angles in the relative energy bin 2.0 -- 2.25 MeV(c),
and at angles beyond  5$^\circ$ in the energy bin 1.25 -- 1.50 MeV(b).
On the other hand, its contribution is inconsequential in the energy
bin 0.5 -- 0.75 MeV (a). This result is in somewhat disagreement 
with that reported in Ref.~\cite{kiku97}, where this component
is reported to be small everywhere below 1.75 MeV relative energy.
Although these authors also perform a quantum mechanical calculation
within DWBA, their treatment of the continuum state is very different
from that of Ref.~\cite{shya99}. Moreover they use a collective
model prescription for
the Coulomb and nuclear form factors, which has a limited
applicability for $^8$B breakup. Bertulani and Gai~\cite{gai98}
have also reported smaller quadrupole component in their analysis of this
data. These authors do not include the nuclear effects in the $E1$
excitations and make use of the eikonal approximation to
calculate the quadrupole nuclear excitation amplitudes. Moreover, the
Coulomb excitation amplitudes have been calculated with the 
PLPA which have been found to be invalid at higher angles (see Fig. 3).
It is also noted in Fig.3 that Coulomb-nuclear interference
effects reduce the $E1$ cross sections at larger angles.

Some authors have studied the importance of the higher order
effects in the Coulomb breakup of $^8$B~\cite{typel94,esbe96,typel97}. 
At RIKEN energies these effects play only a minor role for
this reaction in the kinematical regime of forward angles and low
relative energies~\cite{typel94,esbe96,typel97}. Therefore, the 
conclusions arrived in Ref.~\cite{shya99} about the RIKEN data
are unlikely to be affected 
much by the higher order breakup effects. However, the multi-step
breakup could play an important role at Notre Dame energies, which
is discussed in~\cite{nune99}.  

\begin{figure}[tb]
\begin{center}
\mbox{\epsfig{file=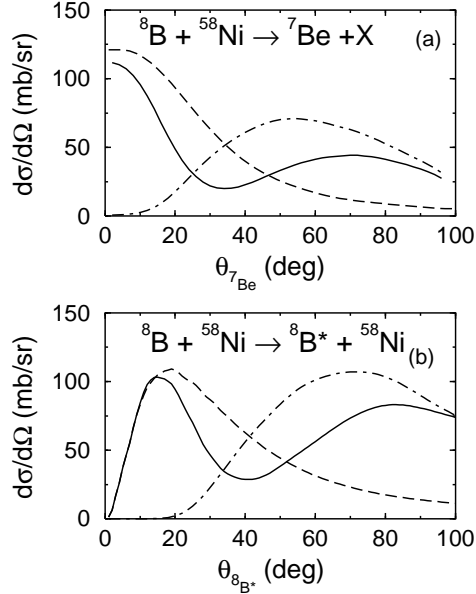,height=8.0cm}}
\end{center}
\caption{{\captionfont
(a) Angular distribution of the $^7$Be fragment emitted in the
breakup reaction of $^8$B on $^{58}$Ni target at the beam energy of 25.8
MeV. The dashed and dashed-dotted lines show the pure Coulomb and
pure nuclear breakup cross sections respectively while their
coherent some is represented by the solid line. (b) Angular distribution
of $^8$B$^*$ in the Coulomb excitation of $^8$B on $^{58}$Ni at
the beam energy of 25.8 MeV. The dashed and dashed-dotted
lines show the cross sections for pure Coulomb and pure nuclear excitation
respectively, while the solid line represents their coherent sum. } }
\label{fig:fige}
\end{figure}

\subsection{$^8$B breakup at subCoulomb energies, Notre Dame data}

In Figs. 5a and 5b, the calculated angular distributions~\cite{shya99}
of $^7$Be and $^8$B$^*$ respectively in a $^8$B induced breakup reaction
on $^{58}$Ni target are shown, at the beam energy of 25.8 MeV.
Pure Coulomb and pure nuclear 
breakup cross sections are represented by the dashed and dashed-dotted 
curves respectively. The ${\it total}$ cross sections 
are represented by the solid lines.
In these calculations also the procedure of single-folding
the respective fragment-target interactions with $^8$B ground and
continuum state wave functions have been used. One can see that  
the angular distributions of $^7$Be and
$^8$B$^*$ are distinctly different from each other. While pure Coulomb and
{\it total} breakup cross sections show a forward peak in case of $^7$Be
(which is typical of the angular distribution of fragments emitted in
breakup reactions), those of $^8$B$^*$ tend to zero as angle goes to zero.
The latter is the manifestation of the adiabatic cut-off typical of the
Coulomb-excitation process. 
In both the cases the nuclear effects are small below 20$^\circ$ and there
is a Coulomb-nuclear interference minimum between 25$^\circ$ - 60$^\circ$.
However the magnitude of various cross sections are smaller in Fig. 5a.
Furthermore, the nuclear-dominated peak occurs at different angles in
Figs 5a ($\simeq$ 55$^\circ$) and 5b ($\simeq$ 70$^\circ$). As discussed 
in~\cite{shya97}, the angles of $^7$Be can be related to those
of $^8$B$^*$. A given $\theta_{7_{Be}}$ gets contributions from a range
of generally larger $\theta_{8_{B^*}}$. This could explain the
shifting of the peaks of various curves to lower angles in Fig. 5a
as compared to the corresponding ones in Fig. 5b. This underlines
the important of three-body kinematics in describing the inclusive
breakup reactions.

The ratio of the experimental integrated breakup cross section of $^7$Be
(obtained by integrating the breakup yields in the angular range,
(45 $\pm$ 6)$^\circ$, of the experimental setup) to Rutherford elastic
scattering of $^8$B is reported to be
(8.1 $\pm$ 0.8$^{+2.0}_{-0.5}$) $\times$ $10^{-3}$~\cite{joha96}. It is
not possible to get this cross section by
directly integrating the angular distributions shown in Fig. 4b 
in this angular range as the corresponding angles belong to $^8$B$^*$
and not to $^7$Be. However, in the three-body case (Fig. 5a), this
can be done in a straight-forward way. This gives a value of 
7.0 $\times$ 10$^{-3}$ which is in close agreement with the
experimental data. Thus, previous failures to explain the 
experimental value may be attributed to the neglect of both the
Coulomb-nuclear interference effects and the three-body
kinematics.

In Fig. 6, the range of the validity of the point-like
projectile approximation (PLPA) and the role of the Coulomb-nuclear 
interference effects on the cross sections of dipole and
quadrupole components is investigated. In Fig. 6a the results for  
pure Coulomb breakup are shown. Dipole and quadrupole components of the cross
section obtained by the single-folding procedure are shown by solid and 
dashed lines respectively, while those obtained with the PLPA 
by solid and dashed lines with solid circles. It can be noted that
PLPA is not valid for angles beyond 20$^\circ$. The condition that
the impact parameter of the collision is larger than the sum of the 
projectile and target radii ($b > R_a + R_t$), 
assumed in applying the Alder-Winther theory, is no longer valid because 
there is a long tail in the $^8$B ground state wave function. We also 
note that the quadrupole component is affected more by the PLPA as compared
to the dipole. The big difference in the dipole and quadrupole cross sections
seen in the PLPA results beyond 20$^\circ$ (where the quadrupole component is
much bigger than the dipole), almost disappears in the
corresponding cross sections obtained by single-folding procedure. 
Nevertheless, the quadrupole cross sections still remain larger than those
of the dipole beyond 30$^\circ$ in the latter case.

In connection with PLPA, it should be made clear that $p$ + target and
the $^7$Be + target potentials {\em do} take into account the finite size of
the $^7$Be and target nuclei. This effect, however, is only important
when two nuclei are very close to each other and is masked by the 
nuclear effects which would be important at those impact parameters.

Dipole and quadrupole cross sections for pure nuclear breakup are shown
in Fig. 6b. The cross sections obtained by summing coherently
the amplitudes of $E1$ and $E2$ components of pure Coulomb and pure 
nuclear breakup are shown in Fig. 6c.
We notice that the Coulomb-nuclear interference effects 
make the contributions of the dipole component of the ${\it total}$ 
cross section larger than those of quadrupole one at all the angles.
This result is quite remarkable as it implies
that the $E2$ component of the total break up cross section in the
$^8$B induced reaction on $^{58}$Ni target is not dominant even
at the subCoulomb beam energies.  Therefore, there is hardly any hope
of determining the $E2$ component of $^8$B breakup by Notre Dame type
of experiment~\cite{joha96}.

This underlines the need for more refined experiments to 
determine the $E2$ component. 
It is clear from Fig. 6c that the  
measurements of the angular distributions may provide  
useful information about the $E2$ component as it is  
different from that of the $E1$ multipolarity.
On the other hand, the angular
distributions of the fragments, calculated within a semiclassical
theory without making the approximation of isotropic angular
distributions in the projectile rest frame, have been shown
to have large $E1$ - $E2$ interference effects~\cite{esbe96}.  
They lead to asymmetries in the momentum distributions of 
the fragments, whose measurements may enable one to put
constraints on the $E2$ component~\cite{davids}. However,
for the better accuracy of this method, improved
calculations including the nuclear effects may be necessary.
 
These results for the nuclear effects in the angular
distribution of $^8$B$^*$ are approximately similar to
those reported in~\cite{vitturi}, where Coulomb and nuclear form
factors are calculated by folding the 
proton-target mean-field (parameterized by a Woods-Saxon function)
by the ground and discretized continuum state $^8$B wave functions.
These authors calculate various cross sections by integrating 
a fixed projectile-target optical potential along a semiclassical
trajectory. However, since the three-body kinematics for the
final state has not been considered by them, a direct 
comparison between their calculations and the data of~\cite{joha96}
is not possible. 

\begin{figure}[tb]
\begin{center}
\mbox{\epsfig{file=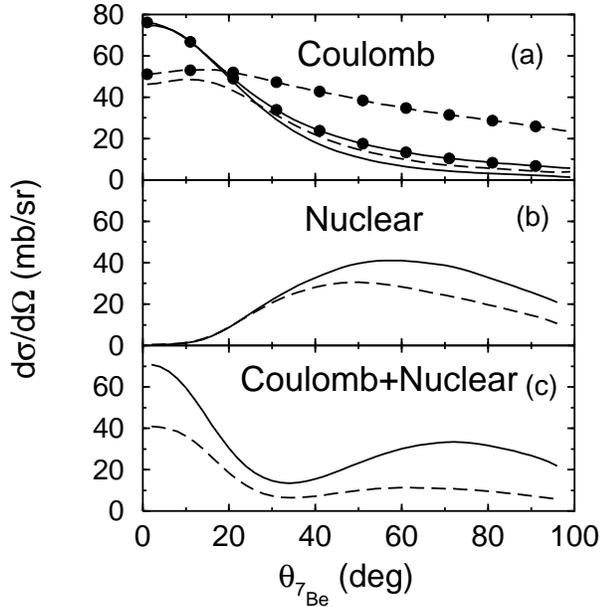,height=8.0cm}}
\end{center}
\caption{{\captionfont
Dipole (solid lines) and quadrupole (dashed lines) components of the
angular distributions of the $^7$Be 
fragment emitted in the breakup reaction of $^8$B on $^{58}$Ni target at
the beam energy of 25.8 MeV. (a) pure Coulomb breakup; also
shown here are the $E1$ (solid lines with solid circles) and $E2$
(dashed lines
with solid circles) cross sections obtained with point-like projectile
and target approximation (Alder-Winther theory), (b) pure
nuclear breakup and (c) Coulomb plus nuclear breakup where the
corresponding amplitudes are coherently summed.}     }
\label{fig:figf}
\end{figure}

\section{Summary and Conclusions}

The Coulomb dissociation method provides a useful tool to 
calculate the cross sections of the difficult-to-measure 
time-reversed processes (ie. radiative capture reactions)
of astrophysical interest. Application of this method
in determining the low-energy cross sections of the 
$^7$Be($p,\gamma)^8$B, which is of considerable interest 
in the context of the Solar Neutrino problem, has yielded
some interesting results since the first pioneering experiment 
performed at RIKEN on the 
$^8$B + $^{208}$Pb $\rightarrow$ $^8$B$^*$ + $^{208}$Pb at 
beam energies around 50 MeV/nucleon. Detailed theoretical
analysis (within the one-step distorted wave Born approximation)
reveal that RIKEN-I and RIKEN-II data are  
almost free from the nuclear effects and are dominated by the $E1$
component for $^7$Be-$p$ relative energies $<$ 0.75 MeV at very
forward angles ($\leq$ 4$^\circ$). The study of the breakup of
$^8$B in this kinematical regime is, therefore, better suited for the
extraction of a reliable $S_{17}(0)$ for the capture reaction
$^7$Be($p,\gamma$)$^8$B at low relative energies.

For the breakup reaction at low energy the Coulomb-nuclear interference
effects are quite important.  
A very striking feature of this
effect is that it makes the $E1$ component of the ${\it total}$ cross
section of the breakup reaction $^8$B + $^{58}$Ni $\rightarrow$ $^7$Be + X
(at the beam energy of 25.8 MeV), larger than the corresponding $E2$
 component 
at all the angles. This renders untenable the main objective of
the Notre Dame experiment of determining the
$E2$ component in the breakup of $^8$B at low beam energies.
The dominance of the $E2$ component for this reaction at this energy,
seen in the semi-classical
Alder-Winther theory of Coulomb excitation has led to this expectation.
However, we note that even in pure Coulomb dissociation process, with 
finite size of the projectile taken into account, the $E2$
components is almost equal to that of $E1$ in the relevant angular range.

It can be said that the feasibility of the Coulomb dissociation
method in determining the $S_{17}(0)$ from the breakup reactions
of $^8$B has been established by identifying the kinematical
regime where the assumptions of this method are well fulfilled.
We now have all the theoretical tools to analyze such experiments,
and a reliable value of $S_{17}$ by means of the Coulomb dissociation
method may be extracted soon. 

\references

\list
 {[\arabic{enumi}]}{\settowidth\labelwidth{[99]}\leftmargin\labelwidth
 \advance\leftmargin\labelsep
 \usecounter{enumi}}
 \def\newblock{\hskip .11em plus .33em minus .07em}
 \sloppy\clubpenalty4000\widowpenalty4000
 \sfcode`\.=1000\relax
 \let\endthebibliography=\endlist

\itemsep=-1pt
\bibitem{bahc89} 
J.N. Bahcall, Neutrino Astrophysics, Cambridge University
Press, New York (1989)

\bibitem{bahc95}
J.N. Bahcall, M.H. Pinsonneault, Rev. Mod. Phys. {\bf 69} (1995) 781.

\bibitem{bahc98} 
J.N. Bahcall, Nucl. Phys. {\bf A 631} (1998) 29.

\bibitem{adel98} 
E.G. Adelberger et al., Rev. Mod. Phys. {\bf 70} (1998) 1265.

\bibitem{baur94} 
G. Baur and H. Rebel, J. Phys.G : Nucl. and Part. Phys. {\bf 20} (1994) 1.

\bibitem{baur97} 
G. Baur and H. Rebel, Ann. Rev. Nuc. Part. Sc. {\bf 46}  (1997) 321.

\bibitem{shya96} 
R. Shyam, I.J. Thompson and A.K. Dutt-Majumder, Phys. Lett. 
{\bf B 371} (1996) 1.

\bibitem{shya97} 
R. Shyam and I.J. Thompson, Phys. Lett. {\bf B 415} (1997) 315.

\bibitem{shya99} 
R. Shyam and I.J. Thompson, Phys. Rev. {\bf C 59} (1999) 2465.  

\bibitem{lang94}
K. Langanke and T.D. Shoppa, Phys. Rev. C {\bf 49} (1994) R1771; {\bf
51} (1995) 2844(E); {\bf 52} (1995) 1709.

\bibitem{gai95}
M. Gai and C.A. Bertulani, Phys. Rev. C {\bf 52} (1995) 1706.

\bibitem{gai98}
C.A. Bertulani and M. Gai, Nucl. Phys. A {\bf 636} (1998) 227.

\bibitem{moto94} 
T. Motobayashi et al., Phys. Rev. Lett. {\bf 73} (1994) 2680.  

\bibitem{kiku97} 
T. Kikuchi et al., Phys. Lett. {\bf B 391} (1997) 261. 

\bibitem{joha96}
Johanes von Schwarzenberg {\it et al.}, Phys. Rev. C {\bf 53} (1996)
R2598.

\bibitem{ald75}
K. Alder and A. Winther, {\it Electromagnetic Excitation}
(North-Holland, Amsterdam, 1975).

\bibitem{alex89} 
A.N.F. Alexio and C.A. Bertulani, Nucl. Phys. {\bf A 505} (1989) 448.
  
\bibitem{type94}
S. Typel and G. Baur, Phys. Rev. C {\ud{50}} (1994) 2104

\bibitem{kimp87}
K.H. Kim, M.H. Park, and B.T. Kim, Phys. Rev.C {\ud{35}} (1987) 363.

\bibitem{esbe96}
H. Esbensen and G.F. Bertsch, Nucl. Phys.A {\bf 600} (1996) 37.
 
\bibitem{nune98a}
F.M. Nunes, R.Shyam and I.J. Thompson, J. Phys.G: Nucl \& Part Phys. 
{\bf 24} (1998) 1575.

\bibitem{nune98b} 
F.M. Nunes and I.J. Thompson, Phys. Rev.  {\bf C 57} (1998) R2818.

\bibitem{nune99} 
F.M. Nunes and I.J. Thompson, Phys. Rev. {\bf C 59} (1999) 2652.

\bm{typel94} S. Typel and G. Baur, Phys. Rev.C {\bf 50}, 2104 (1994).

\bm{esbe95a}  H. Esbensen, G.F. Bertsch and C.A. Bertulani, Nucl. Phys. A
{\bf 581}, 107 (1995).

\bm{typel97} S. Typel, H. Wolter and G. Baur, Nucl. Phys.A {\bf 617},
147 (1997).

\bm{davids} B. Davids et al., Phys. Rev. Lett. {\bf 81}, 2209 (1998).

\bm{vitturi} C.H. Dasso, S.M. Lenzi and A. Vitturi, 
Nucl. Phys. A. {\bf 639} (1999) 635. 

\endlist
\end{document}